\documentclass[3p,times,procedia]{elsarticle}
\usepackage{nupha_ecrc}

\volume{00}

\firstpage{1}

\journalname{Nuclear Physics A}

\runauth{D. Caffarri (CERN) for the ALICE Collaboration}

\jid{nupha}

\jnltitlelogo{Nuclear Physics A}

\usepackage{amssymb}
\usepackage{amsthm}
\usepackage{amsmath}
\usepackage{tabularx}

\usepackage[figuresright]{rotating}
\usepackage{multicol}

\newcommand{\pp}{pp}

\newcommand{\sqrts}{\sqrt{s}}
\newcommand{\sqrtsNN}{\sqrt{s_{\scriptscriptstyle \rm NN}}}

\newcommand{\MeV}{\mathrm{MeV}}
\newcommand{\GeV}{\mathrm{GeV}}
\newcommand{\TeV}{\mathrm{TeV}}

\newcommand{\gev}{\mathrm{GeV}}

\newcommand{\PbPb}{\mbox{Pb--Pb}}
\newcommand{\pPb}{\mbox{p--Pb}}

\newcommand{\pt}{p_{\rm T}}
\newcommand{\constpt}{p_{\rm{T,i}}}
\newcommand{\constsqpt}{p^{2}_{\rm{T,i}}}

\newcommand{\jetpt}{p_{\rm {T, jet}}}

\begin{document}

\begin{frontmatter}



%% Instructions from Editor: Please use the following \dochead only in the preprint version (e-print arXiv etc.); 
%% use empty \dochead{} when submitting to Nuclear Physics A!
\dochead{XXVIth International Conference on Ultrarelativistic Nucleus-Nucleus Collisions\\ (Quark Matter 2017)}
%%\dochead{}
%% e.g. \dochead{17th International Conference on Dynamical Processes in Excited States of Solids}

\title{Exploring jet substructure with jet shapes in ALICE}


\author{D. Caffarri for the ALICE Collaboration}

\address{CERN, Geneva, Switzerland}

\begin{abstract}
The characterization of the jet substructure can give insight into the
microscopic nature of the modification induced on high-momentum partons
by the Quark-Gluon Plasma that is formed in ultra-relativistic
heavy-ion collisions.
Jet shapes allow us to study the modification of
parton to jet fragmentation and virtuality, probing jet energy
redistribution, intra-jet broadening or collimation and possible flavour hierarchy.  
Results of a selected set of jet shapes will be presented for $\pPb$
collisions at $\sqrtsNN = 5.02~\TeV$ and for $\PbPb$ collisions at
$\sqrtsNN = 2.76~\TeV$. Results are also compared with PYTHIA
calculations and models that include in-medium energy loss.
\end{abstract}

\begin{keyword}
jets \sep jet quenching \sep  jet substructure 

\end{keyword}

\end{frontmatter}

% \linenumbers

\section{Introduction}
\label{intro}
The deconfined, highly dense and hot state of nuclear matter created
in $\PbPb$ collisions, known as
Quark-Gluon Plasma, is expected to induce an energy loss of incoming
high-momentum partons, via gluon emission.
This in-medium energy loss modifies the jet yields, the
parton-to-jet fragmentation and the parton virtuality, with respect
to $\pp$ collisions. The measurements of such modifications brings
insight into the mechanisms of energy loss of partons traversing the
medium as well as the possibility to measure the parameters of the medium itself.
Measurements of the same observables in $\pPb$ collisions allow to
study possible cold nuclear matter effects that might affect the
high-$\pt$ particle production and, together with the measurements in
$\pp$ collisions, provide a reference for $\PbPb$ collisions.

Jet shapes are theoretically well defined observables that
allow to study modifications of the fragmentation and virtuality,
exploiting informations on how constituents are distributed in a jet or
considering the clustering history of jets~\cite{Gallicchio:2011xq, Larkoski:2014pca}. 
A selection of jet shapes will be described in this work to probe
different aspects of the possible modifications: the
momentum dispersion ($\pt^{\rm D}$),  the radial moment ($g$), the jet
mass ($M_{\rm {jet}}$) and the $\pt$ distribution of the hard subjet
($z_{g}$).

The momentum dispersion ($\pt^{\rm D}$) defined in Eq.~\ref{eq:shapes}
(left),
quantifies the parton momentum redistribution into jet constituents: jets with fewer and harder
constituents have higher $\pt^{\rm D}$. The radial moment ($g$), defined in
Eq.~\ref{eq:shapes} (center), measures the jet constituents momentum redistribution,
weighted by their distance from the jet axis in the $\eta-\varphi$
plane ($\Delta R_{i}$). 
This shape is sensitive to the collimation or broadening of the jet\footnote{In these definitions,
$\constpt$ refers to the transverse momentum of the constituents of the jets.}.

Due to the subsequent interactions of the incoming high-$\pt$
parton with other partons of the medium, an increase of its
virtuality is expected. This effect would be
observed as an increase of the mass of the jets, once the parton
fragmented~\cite{Majumder:2014gda}. The jet mass is defined as the difference between the energy of the
jet ($E_{\rm jet}$) and its transverse ($\jetpt$) and longitudinal
($p_{\rm {z, jet}}$) momentum, as shown in Eq.~\ref{eq:shapes} (right). 
%\newcolumntype{b}{X}
%\newcolumntype{e}{>{\hsize=.8\hsize}X}
%\newcolumntype{d}{>{\hsize=.7\hsize}X}
%\begin{center}
%\noindent\begin{tabularx}{\textwidth}{deb}
 % \centering
 % \begin{equation} 
  % \pt^{\rm D} = \frac{\sqrt{\sum_{i} \constsqpt}}{\sum_{i}\constpt}  \label{eq:ptd}
  %\end{equation} &
  %\begin{equation}
  % g = \sum_{i}\frac{\constpt}{\jetpt}|\Delta R_{i}|   \label{eq:g}
  %\end{equation} &
  %\begin{equation}
  % M_{\rm {jet}} =\sqrt{E_{\rm jet}^2 - \jetpt^2  - p_{\rm {z, jet}}^2} \label{eq:M}
  %\end{equation}
%\end{tabularx}
%\end{center}
\begin{align}
\label{eq:shapes}
\pt^{\rm D}& = \frac{\sqrt{\sum_{i}
             \constsqpt}}{\sum_{i}\constpt} &
                                     g & =
                                               \sum_{i}\frac{\constpt}{\jetpt}|\Delta
                                               R_{i}|  & 
        M_{\rm {jet} } & =\sqrt{E_{\rm jet}^2 - \jetpt^2  - p_{\rm
                         {z,jet}}^2}
\end{align}
%\label{eq:ptd} 

The momentum distribution between the two hardest subjets is also
considered:~$z_g = \rm{min}(p_{\rm T, 1}, p_{\rm T, 2}) / (p_{\rm T,1} +
p_{\rm T, 2})$, where $p_{\rm T, 1,2}$ indicate the momentum of the two
hardest subjets~\cite{Larkoski:2015lea}. In order to find the these two branches,
the soft radiation is removed from the leading partonic component of the
jet, using the Soft Drop jet grooming algorithm~\cite{Larkoski:2014wba}. 
The measurement of the hardest subjects allows to probe the
role of coherent and de-coherent emitters within one jet in the
medium.

For the characterization of the jet substructure, ALICE focuses on the
low-intermediate transverse momentum ($40<\pt<120~\gev/c$), where
stronger quenching effects are expected but also a larger background due to soft particle production is present.

\section{Jet reconstruction and corrections}

For the $\PbPb$ analyses, the 0-10\% most central collisions were
selected in a sample of data collected during the 2011 LHC Run at
$\sqrts_{NN}=2.76~\TeV$. The $\pPb$ analyses, instead, were performed at
$\sqrts_{NN}=5.02~\TeV$ exploiting a minimum bias and a jet triggered
sample, that was obtained using the ElectroMagnetic CALorimeter
(EMCAL), in order to extend the momentum coverage of
the measurement up to $120~\gev/c$.
Measurements in $\pp$ collisions have also been performed at $\sqrts =
2.76$ and $7~\TeV$ and compared with Monte Carlo
generators~\cite{Cunqueiro:2015dmx}.

In ALICE, jets are reconstructed using the FastJet anti-$k_T$ algorithm with a
resolution parameter $R=0.2$ for the analysis of $\pt^{\rm D}$ and $g$
and $R=0.4$ for the jet mass and $z_g$ analyses. The E-scheme is used
for the recombination and only the charged tracks in $\eta<0.9$ with
$\pt>150~\MeV/c$  are used to reconstruct jets, in order to exploit
the maximum ALICE acceptance in the central rapidity region. 

For $\PbPb$ collisions an event-by-event estimate of the underlying
event momentum and mass densities $\rho$ and $\rho_m$ respectively is
performed using the area based method, implemented in the Fastjet
algorithm~\cite{Cacciari:2007fd}. This average background subtraction is then applied to the
jet shapes, via two different methods: the area derivatives methods~\cite{Soyez:2012hv}
and the constituent subtraction method~\cite{Berta:2014eza}.

In $\pPb$ collisions, the overall background contribution is significantly
smaller than in $\PbPb$ ones but its fluctuations increase due
to event-by-event multiplicity fluctuations. For $\pPb$ analyses,
then, the background was subtracted on average using
unfolding techniques and not subtracted
jet-by-jet~\cite{Acharya:2017goa}.

Residual background fluctuations and detector effect are corrected
using Bayesian two-dimensional unfolding procedure, in order to obtain
fully corrected jet shapes. The procedure uses the RooUnfold
package, using a 4D response matrix that takes into account the jet
$\pt$ and shape at particle and reconstructed levels. For
$\PbPb$ collisions the values are considered at detector level after
correction for the subtraction of the average background and smeared,
to take into account the fluctuations. In order to have this response
matrix, PYTHIA detector level jets have been embedded into $\PbPb$
events and matched with the particle level jets. 
For $\pPb$ collisions, detector level jets were obtained from
embedding detector level four-momentum vectors into $\pPb$ events, in
order not to bias the multiplicity of the event.

\section{Results}
\subsection{Results in $\pPb$ collisions}
\begin{figure}[t]
\begin{center}
\includegraphics[width=0.85\textwidth]{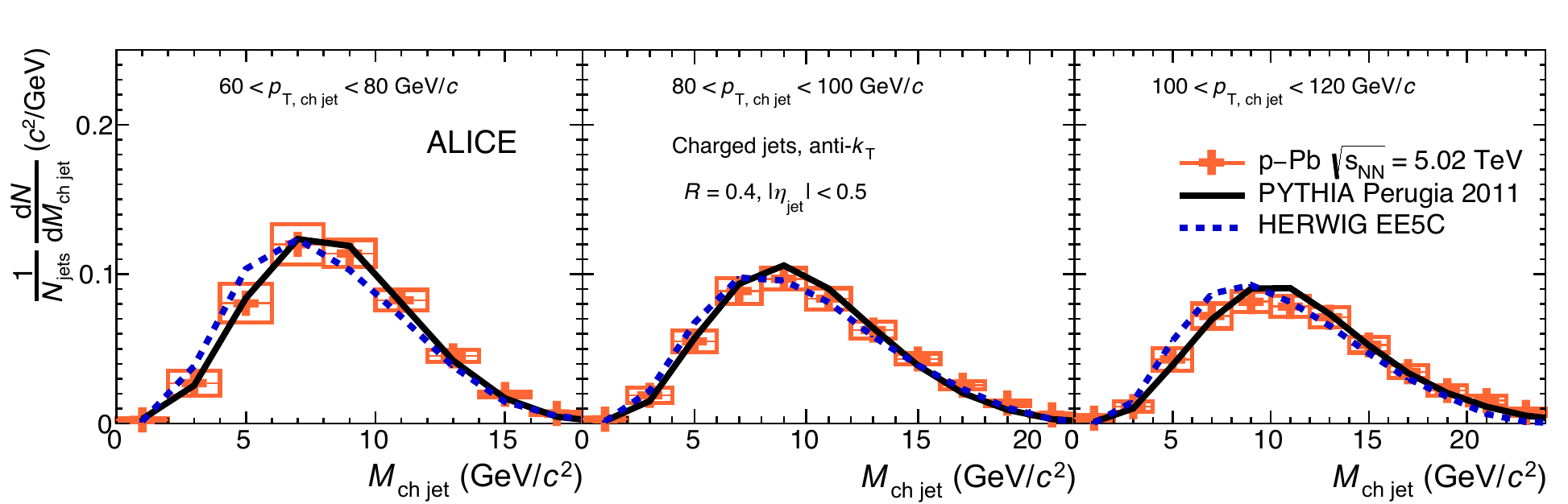}
\includegraphics[width=0.9\textwidth]{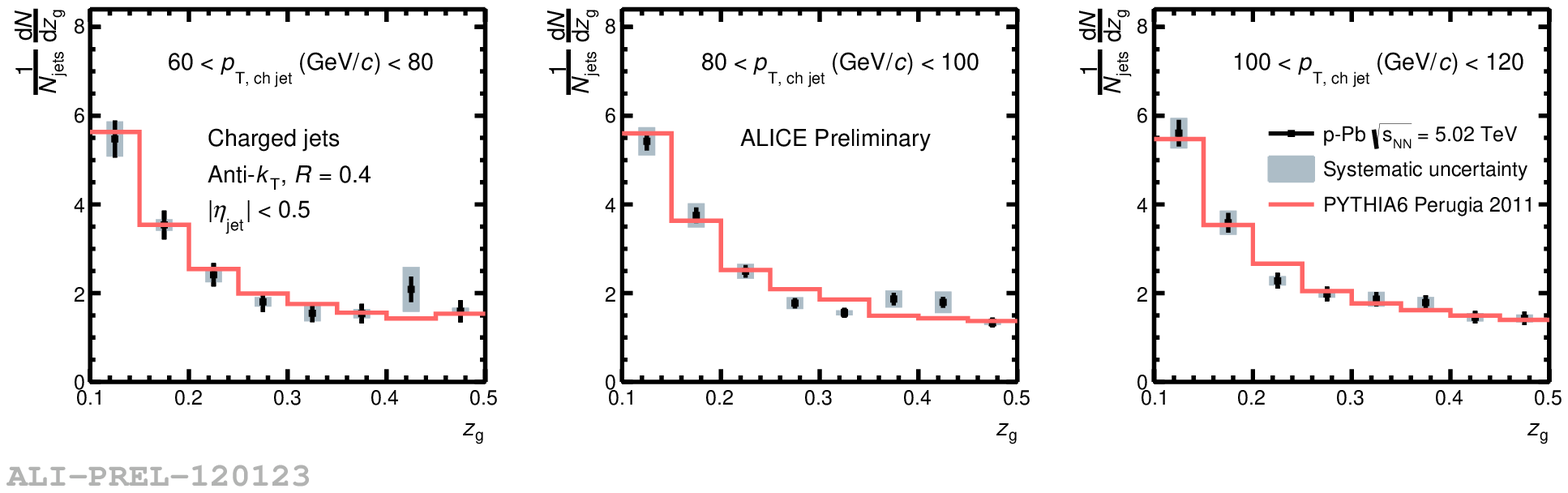}
\end{center}
\caption{ Fully corrected jet mass (top) and $z_g$ (bottom) distributions for anti-$k_{\rm T}$
  jets with $R~=~0.4$ and $60<\pt<120~\gev/c$ in $\pPb$ collisions $\sqrtsNN=5.02~\TeV$, compared with PYTHIA and HERWIG
simulations.}
\label{fig:MpPb}
\end{figure}

Fig.~\ref{fig:MpPb} (top) shows the results of the fully corrected jet mass
distributions measured in $\pPb$ collisions at $\sqrtsNN =
5.02~\TeV$ in three bins of jet transverse momentum between 60 and 120
$\GeV/c$~\cite{Acharya:2017goa}. The measurement is compared with PYTHIA Perugia 11~\cite{Sjostrand:2006za} and
HERWIG~\cite{Bahr:2008pv} Monte Carlo simulations. An agreement within 10-20\% is found
between data and PYTHIA, with some tensions in the tails. Worse
agreement with HERWIG is found, in particular in the low mass
tail. 

Fig.~\ref{fig:MpPb} (bottom) shows the results of the momentum distribution between the two hardest subjets measured in $\pPb$ collisions at $\sqrtsNN =
5.02~\TeV$ in three bins of jet transverse momentum between 60 and 120
$\GeV/c$. The measurement is compared with PYTHIA Perugia 11 and
a good agreement is found. Both these jet
shapes measurements in $\pPb$ collisions can be used as reference measurements for $\PbPb$. 

%\begin{figure}[t]
%\begin{center}
%\includegraphics[width=0.9\textwidth]{2017-Feb-01-zg_unfolded_20GeV_ALL.eps}
%\end{center}
%\caption{Fully corrected $z_g$ distribution for anti-$k_{\rm T}$
 % jets with R=0.4 and $60<\pt<120~\gev/c$ in $\pPb$ collisions, compared with PYTHIA Perugia 11.}
%\label{fig:zgpPb}
%\end{figure}

\subsection{Results in $\PbPb$ collisions}

\begin{figure}[t]
\begin{center}
\includegraphics[width=0.85\textwidth]{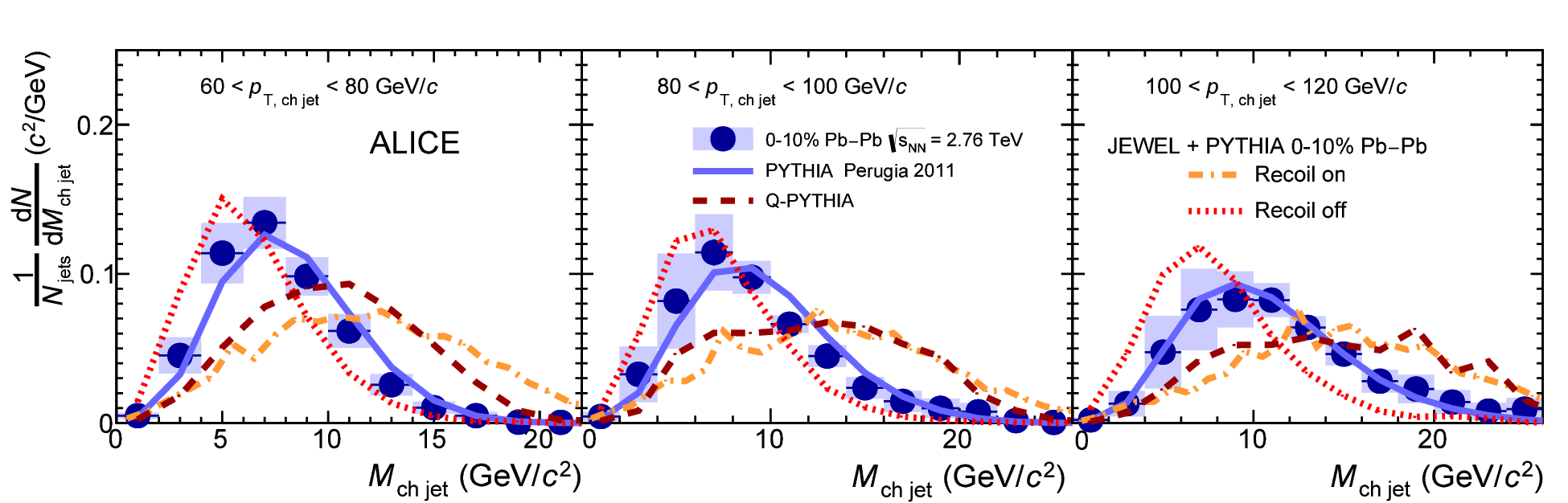}
\end{center}
\caption{Fully corrected jet mass distribution for anti-$k_{\rm T}$
  jets with $R~=~0.4$ and $60<\pt<120~\gev/c$ in $\PbPb$ collisions $\sqrtsNN=2.76~\TeV$, compared with PYTHIA Perugia
  11, Q-PYTHIA and JEWEL models.}
\label{fig:MPbPbModels}
\end{figure}

Fig.~\ref{fig:MPbPbModels} shows the results of the fully corrected jet mass
distributions measured in $\PbPb$ collisions at $\sqrtsNN =
2.76~\TeV$ in three bins of jet transverse momentum between 60 and 120
$\GeV/c$~\cite{Acharya:2017goa}. This measurement shows a hint of a shift towards smaller jet
mass values with respect to the $\pPb$ case for $\pt < 100~\gev/c$. 
In order to take into account the different
quark and gluons components and the different shape in the underlying
jet-$\pt$ spectrum, a ratio of the jet mass
distributions is considered and compared with PYTHIA pp collisions at
the two energies. A hint of difference is observed also between the
two ratios.  
A $1\sigma$ difference is observed when considering the mean jet mass
for $60<\pt<80~\gev/c$.

Fig.~\ref{fig:MPbPbModels} shows also the comparison of the measurements with
different theoretical model calculations. 
Data lie between PYTHIA Perugia 11 and JEWEL~\cite{Zapp:2013vla}
in the case when recoil partons do not contribute to the
final state hadrons.
Q-PYTHIA~\cite{Armesto:2009fj} and JEWEL, when
including the recoil process, predict a too large jet masses. 

\begin{figure}[t]
\begin{center}
\includegraphics[width=0.4\textwidth]{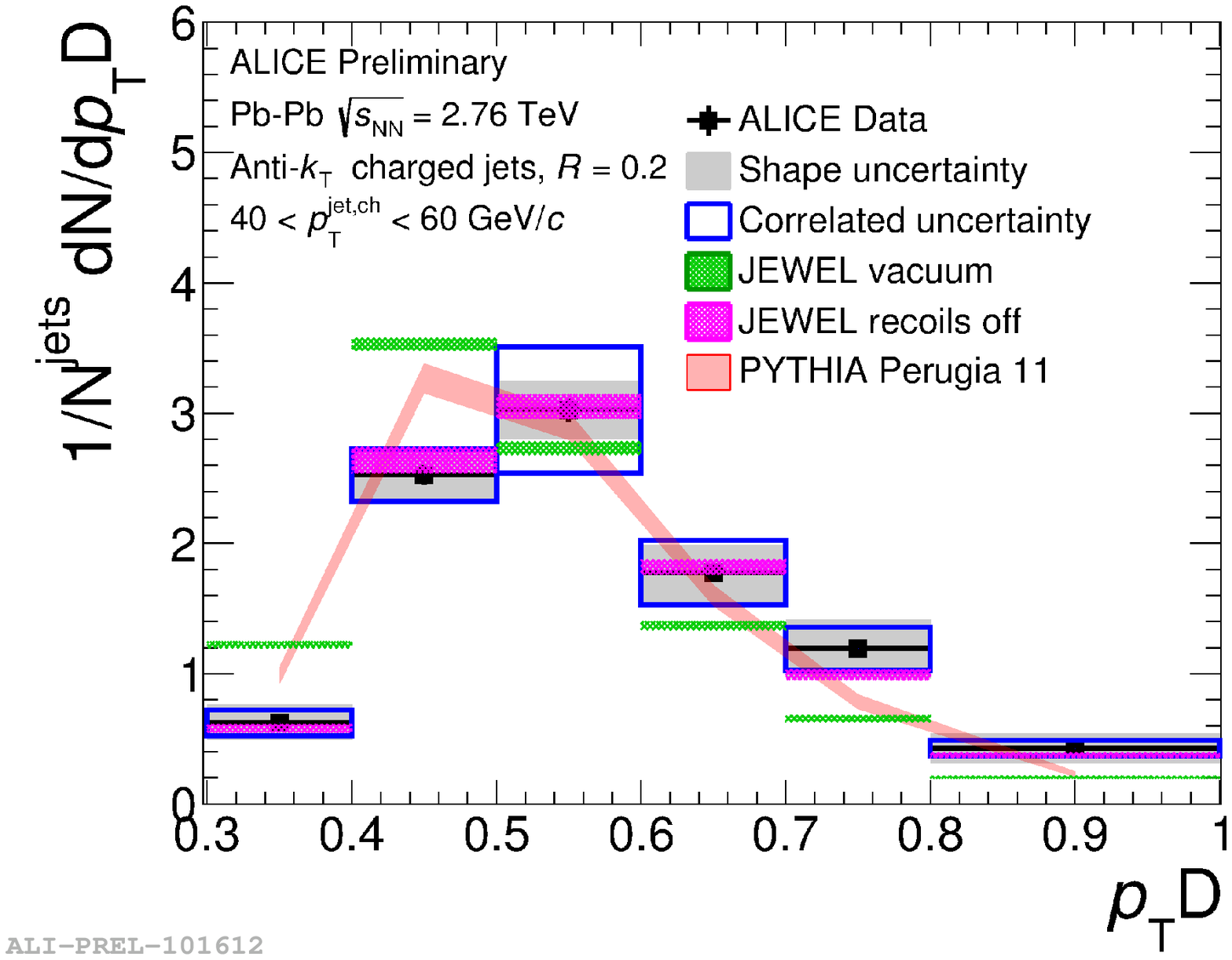}
\includegraphics[width=0.4\textwidth]{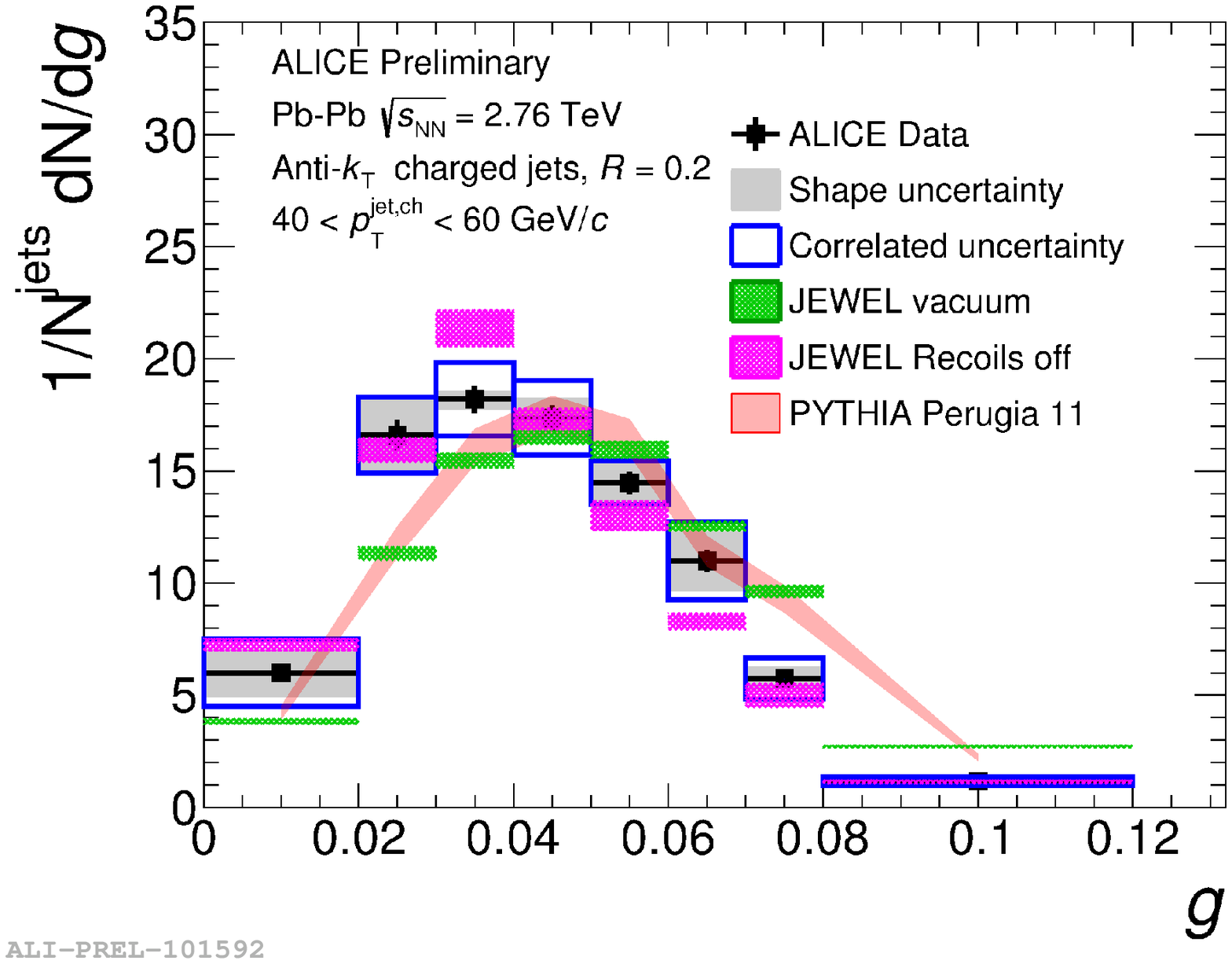}
\end{center}
\caption{Fully corrected $\pt^{\rm D}$ (left) and $g$ (right) distributions for anti-$k_{\rm T}$
  jets with $R~=~0.2$ and $40<\pt<60~\gev/c$ in $\PbPb$ collisions at $\sqrtsNN=2.76~\TeV$, compared with PYTHIA Perugia
  11 and JEWEL models.}
\label{fig:ShapesPbPbModels}
\end{figure}

Fig.~\ref{fig:ShapesPbPbModels} shows the results of the fully
corrected $\pt^{\rm D}$ (left) and $g$ (right), measured in $\PbPb$ collisions at $\sqrtsNN =
2.76~\TeV$ for jets with $40<\pt < 60~\gev/c$. 
Results are compared with PYTHIA Perugia 11.  The momentum dispersion
distribution is shifted to higher values in the $\PbPb$ measurement
with respect to the $\pp$ Monte Carlo. The radial moment distribution
is shifted lo lower values in $\PbPb$ collisions with respect to
PYTHIA. 
In Fig.~\ref{fig:ShapesPbPbModels}, results are also compared with
JEWEL with both options of medium-jet recoil interaction and they are better
described in the case when this option is switched off. 
The underlying physics mechanism in JEWEL model is based on the fact
that soft modes are transported al large angles
relative to the jet axis and this leads to a collimation of the jet. 

All the ALICE jet shapes measurements show a consistent
picture compatible with jets more collimated and with a harder
fragmentation, in $\PbPb$ collisions than for the $\pp$ case. 

%% References


\bibliographystyle{elsarticle-num}
\bibliography{<your-bib-database>}



\end{document}